\documentstyle[epsf,aps]{revtex}
\title{\bf Non-Abelian Weizs\"{a}cker-Williams field and a
two-dimensional effective color charge density for a very large
nucleus.}

\author{Yuri V. Kovchegov}

\begin{document}
\maketitle
\begin{center}
 Physics Department, Columbia University, New York, NY 10027, USA \\
\ \\
 
\end{center}
\begin{abstract}

   We consider a very large ultra-relativistic nucleus. Assuming a
simple model of the nucleus and weak coupling we find a classical
solution for the gluon field of the nucleus and construct the
two-dimensional color charge density for McLerran-Venugopalan model
out of it. We prove that the density of states distribution, as a
function of color charge density, is Gaussian, confirming the
assumption made by McLerran and Venugopalan.

\ \\
PACS number(s): 12.38.Bx, 24.85.+p
\end{abstract}


\section{Introduction.}

   Consider a very large nucleus, probably larger than can be
physically realized. The nucleons are distributed homogeneously inside
the nucleus. Recently L. McLerran and R. Venugopalan proposed a
program of computing the gluon distribution fuctions for such a
nucleus at small $x$ [1,2].

   One of the interesting problems in the McLerran-Venugopalan model
for the small-$x$ part of the gluon distribution of such a large
nucleus [1,2] is finding the classical solution for the gluon field
treating the valence quarks of the nucleons in the nucleus as
recoiless sources, which are delta functions along the light-cone when
the nucleus is moving near the velocity of light. A convenient way to
deal with the problem is by working in the light-cone gauge. The
source is characterized by a two-dimensional color charge density
$\rho(\underline{x}) $, where $\underline{x}$ is a vector in the
transverse direction.  The proposed model assumes that in order to
find the average value of any observable having longitudinal coherence
length long compared to the nucleus, one calculates this observable
for a given $\rho(\underline{x})$, and then averages it over all $\rho
$ with the measure
\begin{eqnarray}
 \int [d \rho] \exp\left( -{1\over 2{\mu}^2} \int d^2 {x} {\rho ^2} (
\underline{x}) \right),
\label{distrib}
\end{eqnarray}
where ${\mu}^2$ is the average charge density squared.

   We consider a large nucleus consisting of ``nucleons'', which for
simplicity of description are chosen to be just quark-antiquark pairs
(see Fig. 1). Valence quark and antiquark are treated as point
particles free to move inside of the nucleon, but unable to get out.
\begin{figure}
\begin{center}
\epsfxsize=6cm
\leavevmode
\hbox{ \epsffile{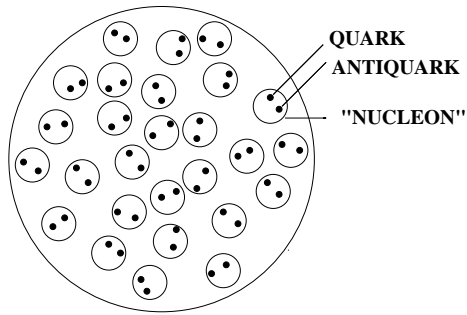}}
\caption{Nucleus with ``nucleons'' being quark-antiquark pairs.}
\end{center}
\end{figure}
We are interested in the gluon field of the ultrarelativistic nucleus
viewed in the laboratory frame. We assume that the field in each
individual nucleon is not large. This allows us to approximate the
covariant gauge potential of each quark by a single gluon exchange. In
a covariant gauge the classical field of a single ultrarelativistic
particle is proportional to a delta-function in the $ x_{-} $
direction [3]. Since in our model of ultrarelativistic nucleus
different quarks have different $ x_{-} $ coordinates in the
laboratory frame, the fields of individual quarks don't overlap, which
allows us to superimpose them and justifies our single gluon exchange
approximation. Then the total field of the nucleus in the covariant
gauge is the sum of the quark fields. We make a gauge transformation
which changes the total potential to the light-cone gauge. So, we get
a solution to the classical equations of motion in light-cone gauge,
where the field, $A_\mu$, is directly related to the gluon
distribution in the small-$x$ region [1].

      Following McLerran and Venugopalan the nucleus is considered to
be very large, thus although the field of each individual nucleon is
weak, the total field is strong at low momentum in the light-cone
gauge due to the overlap of the fields of a huge number of
nucleons. Still we can neglect the contributions of several nucleons,
without changing the answer, i.e., we work in the leading power of the
number of nucleons.

       We may treat the source as classical only when we are at
sufficiently small momenta that the individual quarks can not be
resolved. It was shown in [1] that this requires that ${\underline
{k}}^2 \ll \mu^2$, where $\underline {k}$ is the typical momentum
scale. The weak coupling approximation is valid when ${\underline
{k}}^2 \gg \Lambda_{QCD}^2$. Then the momentum range we consider is
$\Lambda_{QCD}^2 \ll {\underline {k}}^2 \ll \mu^2$.

       Now the task is to construct the two-dimensional charge
density, giving the correct classical solution. This is done by just
substituting the classical solution in the equations of motion. The
density we find this way happens to satisfy the Gaussian
distribution. We show this by calculating the correlation functions of
the densities at different transverse points and proving that they are
exactly what one would expect for the Gaussian distribution
(\ref{distrib}). That is we justify the method for averaging the
observables proposed by McLerran and Venugopalan.

   In sect. 2 we calculate the solution of the classical equations of
motion - the non-Abelian Weizs\"{a}cker-Williams field.

   In sect. 3 we construct the two-dimensional charge density.

   In sect. 4 we show that the charge density has a Gaussian
distribution by calculating the correlation functions.

   In sect. 5 we confirm the techniques proposed in [1].

\section{Approximate solution.}

  We start with some random distribution of nucleons in the nucleus
  and quarks and antiquarks in the nucleons. The nucleons , in the
  rest frame of the nucleus, are assumed to be spheres with equal
  radius, and the quarks and antiquarks are distributed randomly
  inside each sphere, with equal probability to be at any place inside
  the sphere, but with zero probability to get outside.  The density
  in the rest frame is given by
\begin{eqnarray}
 \rho (\vec {x}) = \sum_{a=1}^8 T^a {\rho}^a (\vec {x})
\end{eqnarray}
with
\begin{eqnarray}
 {\rho}^a (\vec {x})= g \sum_{i=1}^N (T_i^a)
[\delta({\vec{x}}-{\vec{x}}_i )-\delta({\vec{x}}-{\vec{x}}'_i ) ] ,
\end{eqnarray}
where $\vec{x}_i$ is the coordinate of a quark in the ith nucleon
(there are $N$ nucleons in the nucleus), $\vec{x}'_i$ is the
coordinate of the antiquark, $T^a$ are generators of $ SU(3) $ in
color space, $(T_i^a) $ are similar generators in the color space of
each nucleon. The reason we separate them is because $\rho ^a $ comes
from the current $j^a_{\mu}= g {\overline q}_\alpha \gamma_{\mu}
(T^a)_{\alpha \beta} q_\beta $, so the expression for $\rho ^a $
should include a $T^a$ acting in each individual nucleon's color
space.

     In the laboratory frame for the ultra-relativistic nucleus the
density is
\begin{eqnarray}
 {\rho} (\underline {x}, x_{-})= {g \over {\sqrt {2}}} \sum_{a=1}^8
\sum_{i=1}^N T^a (T_i^a)
[\delta(x_{-}-x_{-i})\delta({\underline{x}}-{\underline{x}}_i
)-\delta(x_{-}-x'_{-i})\delta({\underline{x}}-{\underline{x}}'_i ) ] .
\end{eqnarray}
Assuming that the coupling is weak and using the expression for the
potential of a single particle in a covariant gauge [3, Appendix A] we
approximate the field of the nucleus by superposition as
\begin{eqnarray}
A'_{+} = {-} { g \over {2 \pi }} \sum_{a=1}^8 \sum_{i=1}^N
 T^a (T_i^a) \left[ \delta(x_{-}-x_{-i}) \ln
 (|{\underline{x}}-{\underline{x}}_i| \lambda ) \right. \nonumber\\
 \left.  -\delta(x_{-}-x'_{-i}) \ln
 (|{\underline{x}}-{\underline{x}}'_i| \lambda ) \right] , \underline
 A' =0 , A'_{-}=0 ,
\label{covfd}
\end{eqnarray}
where $\lambda$ is some infrared cutoff. The prime at the field or the
 field strength denotes the covariant gauge. The field strength in the
 covariant gauge is then 
\begin{eqnarray} F'_{+\perp}={ g \over {2 \pi}}
 \sum_{a=1}^8 \sum_{i=1}^N T^a (T_i^a) \left(
 \delta(x_{-}-x_{-i}){{\underline{x}}-{\underline{x}}_i \over
 |{\underline{x}}-{\underline{x}}_i|^2}
 -\delta(x_{-}-x'_{-i}){{\underline{x}}-{\underline{x}'}_i \over
 |{\underline{x}}-{\underline{x}'}_i|^2} \right) .
\label{covfs}
\end{eqnarray}
From here on the subscript $\perp$ will mean that the object is a
vector in the transverse space over this index.

   We now do a gauge transformation to transorm this field into the
light-cone gauge. The potential in a new gauge is
\begin{eqnarray}
A_{\mu} = S A'_{\mu} S^{-1} - { i \over g } (\partial _{\mu} S)
S^{-1}.
\end{eqnarray}
Requiring the new gauge to be the light-cone gauge, $A_+ =0 $, we
obtain
\begin{eqnarray}
S(\underline {x}, x_{-} ) = \mbox{P} \exp 
\left( { -ig \int_{-\infty}^{x_{-}} dx'_{-} A'_{+} 
(\underline {x}, x'_{-})} \right).
\label{smatr}
\end{eqnarray}
Then the field in the light cone gauge is
\begin{eqnarray}
{\underline {A}}(\underline {x}, x_{-} ) = \int_{-\infty}^{x_{-}}
dx'_{-} F_{+\perp}(\underline {x}, x'_{-} ) = \int_{-\infty}^{x_{-}}
dx'_{-} S(\underline {x}, x'_{-} ) F'_{+\perp}(\underline {x}, x'_{-}
) S^{-1} (\underline {x}, x'_{-} ) ,
\end{eqnarray}
Only transverse components are non-zero.  Substituting
$F'_{+\perp}(\underline {x}, x'_{-} )$ from (\ref{covfs}) we get
\begin{eqnarray}
{\underline {A}} (\underline {x}, x_{-} ) = { g \over {2 \pi
}} \sum_{a=1}^8 \sum_{i=1}^N (T_i^a) \left( S(\underline {x},
x_{-i} ) T^a S^{-1} (\underline {x},
x_{-i}){{\underline{x}}-{\underline{x}}_i \over
|{\underline{x}}-{\underline{x}}_i|^2}\theta (x_{-} - x_{-i})
\right. \nonumber
\end{eqnarray}
\begin{eqnarray}
 - \left. S(\underline {x}, x'_{-i} ) T^a S^{-1} (\underline {x},
x'_{-i}){{\underline{x}}-{\underline{x}'}_i \over
|{\underline{x}}-{\underline{x}'}_i|^2}\theta (x_{-} -
x'_{-i})\right).
\label{clsol}
\end{eqnarray}
 This is our estimate of the solution of the classical equations of
motion for a given configuration of quarks inside the nucleons and
nucleons inside the nucleus. Formula [\ref{clsol}] gives us the
non-Abelian Weizs\"{a}cker-Williams field generated by the valence
quarks.

\section{Two-dimensional color charge density.}

   The equation of motion is
\begin{eqnarray}
 D_\mu F^{\mu \nu} = J^\nu .
\label{clas1}
\end{eqnarray}
In the McLerran-Venugopalan model [1] the classical current $J^\mu $
only has components in the + direction and is proportional to a delta
function of $x_{-}$:
\begin{eqnarray}
J^\mu (x ) = \delta ^{\mu +} \delta (x_{-}) \rho (\underline {x}).
\label{clas2}
\end{eqnarray}   
This can be treated as a definition of the two-dimensional color
density. Our goal now is to construct $\rho (\underline {x})
$. Integrating both sides of (\ref{clas1}) over $x_{-}$ and using
(\ref{clas2}) gives
\begin{eqnarray}
\rho (\underline {x}) & = & \int_{-\infty}^{+\infty} dx_{-} D_i F_{+i}
(\underline {x}, x_{-} ) \nonumber\\ & & = \int_{-\infty}^{+\infty}
dx_{-} \{ \partial_i F_{+i}(\underline {x}, x_{-} ) - i g [A_i
(\underline {x}, x_{-} ), F_{+i} (\underline {x}, x_{-} )] \} ,
\label{dens1}
\end{eqnarray}
where $i=1,2$ (transverse direction).  Using ${\underline {A}}
(\underline {x}, x_{-} )$ from (\ref{clsol}) , we can calculate
$F_{+\perp} (\underline {x}, x_{-} )$. Substituting both in
(\ref{dens1}) we end up with the following expression for the density:
\begin{eqnarray}
 \rho (\underline {x}) = { g } \sum_{a=1}^8
\sum_{i=1}^N (T_i^a)[S(\underline {x}_i, x_{-i} ) T^a S^{-1}
(\underline {x}_i, x_{-i}) \delta (\underline {x} - \underline {x}_i)
\nonumber\\ - S(\underline {x}'_i, x'_{-i} ) T^a S^{-1} (\underline
{x}'_i, x'_{-i}) \delta (\underline {x} - \underline {x}'_i)].
\label{dens2}
\end{eqnarray}
The details of calculations are presented in Appendix B. We can see
now that our expression for two-dimensional density is just a rotation
of the three-dimensional density (in the laboratory frame) we started
with:
\begin{eqnarray}
\rho (\underline {x}) = \int_{-\infty}^{+\infty} dx_{-} S(\underline
{x}, x_{-}) {\sqrt {2}}\rho (\underline x, x_{-}) S^{-1} (\underline 
{x}, x_{-}).
\end{eqnarray}

     In the expression for the light-cone potential we had an infrared
cutoff $\lambda$, so it may seem that this cutoff will appear in the
$S(\underline {x}, x_{-} )$, and, consequently in $\rho (\underline
{x})$. However this is not the case, because, although the light-cone
potential is cutoff dependent, $S(\underline {x}, x_{-} )$ is not.  To
see this let's do an explicit calculation of $S(\underline {x}, x_{-}
)$: substituting (\ref{covfd}) into (\ref{smatr}) and using the
definition of the path-ordered exponential we obtain
\begin{eqnarray}
S(\underline {x}, x_{-} ) = \prod_{i=1}^N \left( \theta (x'_{-i} -
x_{-i}) e^{ \Sigma_i } e^{\Sigma'_i } + \theta (x_{-i} - x'_{-i})
e^{\Sigma'_i} e^{\Sigma_i} \right),
\end{eqnarray}
with 
\begin{eqnarray}
\Sigma_i = {i g^2 \over {2 \pi }} \sum_{a=1}^8 T^a (T_i^a) \ln
(|\underline {x} - {\underline {x}}_i| \lambda) \theta (x_{-} -
x_{-i}), \nonumber
\end{eqnarray}
\begin{eqnarray}
\Sigma'_i = - {i g^2 \over {2 \pi}} \sum_{a=1}^8 T^a (T_i^a)
\ln(|\underline {x} - {\underline {x'}}_i| \lambda)\theta (x_{-} -
x'_{-i}).
\end{eqnarray}
But the matrices we exponentiate ($\Sigma_i$ and $\Sigma'_i$) commute
(for the same $i$), so
\begin{eqnarray}
S(\underline {x}, x_{-} ) = \prod_{i=1}^N 
\exp \left[ {i g^2 \over {2 \pi }}
\sum_{a=1}^8 T^a (T_i^a) \ln \left({|\underline {x} - {\underline
{x}}_i| \over { |\underline {x} - {\underline {x}}'_i|}} \right)
\theta (x_{-} - x_{-i})\right].
\label{cutoff}
\end{eqnarray}
Here we neglected the contribution of the ``last'' nucleon, i.e. the
nucleon (or several nucleons) whose quarks or antiquarks may overlap
the point $x_{-}$ at which we calculate $S(\underline {x}, x_{-}
)$. These nucleons may potentially cause us some trouble, but the
philosophy of the large nucleus approximation implies that the fields
of individual nucleons are small, and we construct a strong field out
of a large number of nucleons. The contribution of each individual
nucleon is negligible, it's their sum which matters. That means we can
neglect these ``last'' nucleons.

    Another way to say this is that we want to do a calculation
keeping the leading powers of $N$ only. Then dropping few nucleons
will not change our result.

    From (\ref{cutoff}) we see that $S(\underline {x}, x_{-} )$ is
cutoff independent, and so is the density $\rho (\underline {x})$.

\section{Calculation of correlation functions.}

 Now that we found the charge density, let's show that it's
distribution is Gaussian by calculating the density correlation
functions. First we note that the average density is zero, as
expected: $< \rho^a (\underline {x}) > =0$, where $< \ldots >$ denote
the averaging over all possible positions of quarks and antiquarks in
the nucleons, and nucleons in the nucleus, as well as averaging over
all possible colors (keeping each nucleon color neutral).

     For two densities correlation function we have :
\begin{eqnarray}
<\rho^a (\underline {x})\rho^b (\underline {y})> = \prod_{k=1}^N \int
{ d^3 r_k \over {(4/3) \pi R^3}} { d^3 x_k d^3 x'_k \over {[(4/3) \pi
a^3]^2}} (\alpha {\overline \alpha}| \rho^a (\underline {x})\rho^b
(\underline {y}) |\beta {\overline \beta}),
\end{eqnarray}
where $R$ is the radius of the nucleus, $a$ is the radius of the
nucleons, $r_k$ is the position of the center of the $k$th nucleon in
the nucleus (in the rest frame), $x_k$ and $x'_k$ are positions of the
quarks in the nucleons; $(\alpha {\overline \alpha}| \ldots |\beta
{\overline \beta})$ implies an average over all color-neutral states
of the nucleons.

      Using (\ref{dens2}) we obtain (since $\rho^a (\underline {x}) =
2 \mbox {Tr} [T^a \rho (\underline {x})]$)
\begin{eqnarray}
<\rho^a (\underline {x})\rho^b (\underline {y})> = { g^2 }
\prod_{k=1}^N \int { d^3 r_k \over {(4/3) \pi R^3}} { d^3 x_k d^3 x'_k
\over {[(4/3) \pi a^3]^2}} (\alpha {\overline \alpha}| \sum_{c,d=1}^8
\sum_{i,j=1}^N (T^c_i) (T^d_j) \nonumber
\end{eqnarray}
\begin{eqnarray}
\times \left\{ 2 \mbox {Tr} [ T^a S(\underline {x}_i, x_{-i} ) T^c
S^{-1} (\underline {x}_i, x_{-i})] \delta (\underline {x} - \underline
{x}_i) - 2 \mbox {Tr} [ T^a S(\underline {x}'_i, x'_{-i} ) T^c S^{-1}
(\underline {x}'_i, x'_{-i})] \delta (\underline {x} - \underline
{x}'_i) \right\} \nonumber
\end{eqnarray}
\begin{eqnarray}
\times \left\{ 2 \mbox {Tr} [ T^b S(\underline {x}_j, x_{-j} ) T^d
S^{-1} (\underline {x}_j, x_{-j})] \delta (\underline {y} - \underline
{x}_j) - 2 \mbox {Tr} [ T^b S(\underline {x}'_j, x'_{-j} ) T^d S^{-1}
(\underline {x}'_j, x'_{-j})] \delta (\underline {y} - \underline
{x}'_j) \right\} |\beta {\overline \beta}).
\label{corr1}
\end{eqnarray}
In $S(\underline {x}_i, x_{-i} )$ the ``last'' nucleon is the $i$th
nucleon. Applying the same arguments we had before we can drop this
``last'' nucleon. Then there will be no $(T^c_i)$ matrices in
$S(\underline {x}_i, x_{-i} )$ and $S(\underline {x}'_i, x'_{-i}
)$. It is convenient to label nucleons according to their coordinates
along the $x_{-}$ axis. The greater the coordinate, the greater the
number of the nucleon. Without any loss of generality we can assume
that $i \ge j$. Then if $i>j$ the $(T^c_i)$ matrix in front will be
the only matrix in the $i$th nucleon color space for the term
corresponding to fixed $i$ and $j$ in our expression. So, when we do
the color averaging it will give zero [$ \mbox {Tr} (T^c_i) =0$].
That means that $i=j$. Then color averaging in the $i$th nucleon space
gives ${1 \over N_c} \mbox {Tr} [(T^c_i)(T^d_i)]= {1 \over {2 N_c}}
\delta^{cd}$.

    Each density is a difference of the quark and antiquark parts. In
(\ref{corr1}) one can easily see that the product of quark (and
antiquark) components gives a delta-function of $\underline {x}$ and
$\underline {y}$ , while the product of quark and antiquark components
gives some smooth function. In most of the physical applications we'll
be looking at the scales much smaller than the nucleon's radius $a$
[4]. At these scales we can neglect this cross terms with respect to
the delta-function terms:
\begin{eqnarray}
<\rho^a (\underline {x})\rho^b (\underline {y})> = { g^2 \over {
N_c}} {1 \over {\pi a^2}} \delta (\underline {x} - \underline {y})
\prod_{l=1}^N \int { d^3 r_l \over {(4/3) \pi R^3}} \sum_{i=1}^N {3
\over 2} \sqrt {1-{(\underline {x} - {\underline {r}}_i)^2 \over a^2}}
\prod_{k=1}^{i-1} \int { d^3 x_k d^3 x'_k \over {[(4/3) \pi a^3]^2}}
\nonumber\\ \times (\alpha {\overline \alpha}| \sum_{c=1}^8 4 \mbox
{Tr} [ T^a S(\underline {x}_i, x_{-i} ) T^c S^{-1} (\underline {x}_i,
x_{-i})] \mbox {Tr} [ T^b S(\underline {x}_i, x_{-i} ) T^c S^{-1}
(\underline {x}_i, x_{-i})] |\beta {\overline \beta}).
\end{eqnarray}
 After dropping the ``last'' nucleon $S(\underline {x}, x_{-i} )=
S(\underline {x}, x'_{-i} )$.
 
    For two traceless $3 \times 3$ matrices $M$ and $N$ the following
formula is true:
\begin{eqnarray}
\sum_{a=1}^8 \mbox {Tr} [M T^a ] \mbox {Tr} [N T^a ] = {1 \over 2}
\mbox {Tr} [MN].
\end{eqnarray}
 Using this, and making some approximation when integrating over $r_i$
($a \ll R$), we get
\begin{eqnarray}
<\rho^a (\underline {x})\rho^b (\underline {y})> = \delta^{ab} { 3 g^2
\over {2 N_c}} { N \over {\pi R^2}} \delta (\underline {x} -
\underline {y}) \sqrt {1 - {{\underline {x}}^2 \over R^2}}.
\label{corr4}
\end{eqnarray}
Here , for the first time we made an assumption about the geometry of
the nucleus and nucleons in it - when doing the average over the
positions of quarks and nucleons we assumed that the nucleus and
nucleons are spherical in the rest frame. For a cylindrical nucleus
(in $z$-direction) one would get $<\rho^a (\underline {x})\rho^b
(\underline {y})> = \delta^{ab} { g^2 \over { N_c}} { N \over {\pi
R^2}} \delta (\underline {x} - \underline {y})$.

     By employing a similar techniques of dropping the ``last''
nucleon in the $S(\underline {x}_i, x_{-i} )$ and keeping the leading
powers of $N$ only, we can show that the four-density correlation
function is
\begin{eqnarray}
<\rho^a (\underline {x})\rho^b (\underline {y})\rho^c (\underline
{z})\rho^d (\underline {w})> = <\rho^a (\underline {x})\rho^b
(\underline {y})><\rho^c (\underline {z})\rho^d (\underline {w})> +
<\rho^a (\underline {x}) \rho^c (\underline {z})> \nonumber
\end{eqnarray}
\begin{eqnarray}
\times < \rho^b (\underline {y}) \rho^d (\underline {w})> + <\rho^a
(\underline {x}) \rho^d (\underline {w})><\rho^b (\underline {y})
\rho^c (\underline {z})>
\end{eqnarray}
and prove a similar formula for a correlation function with any even
number of densities, i.e., that it can be represented as
\begin{eqnarray}
<\rho^{a_1} (\underline {x}^1) \ldots \rho^{a_{2n}} (\underline
{x}^{2n})> = <\rho^{a_1} (\underline {x}^1) \rho^{a_2} (\underline
{x}^2)> \ldots <\rho^{a_{2n-1}} (\underline {x}^{2n-1}) \rho^{a_{2n}}
(\underline {x}^{2n})> \nonumber
\end{eqnarray}
\begin{eqnarray}
+\ \mbox {permutations} ,
\label{corr2}
\end{eqnarray}
where $\underline {x}^1, \ldots , \underline {x}^{2n}$ are just $2n$
arbitrary points in the nucleus.  Also we can show that a correlation
function with an odd number of densities is zero to all orders in $N$
(for details see Appendix C):
\begin{eqnarray}
<\rho^{a_1} (\underline {x}^1) \ldots \rho^{a_{2n+1}} (\underline
{x}^{2n+1})> =0.
\label{corr3}
\end{eqnarray}

\section{Conclusions.}

By calculating the two densities correlation function (\ref{corr4})
and proving (\ref{corr2}) and (\ref{corr3}) we showed that for any $n$
points $\underline {x}^1 \ldots \underline {x}^{n}$ :
\begin{eqnarray}
<\rho^{a_1} (\underline {x}^1) \ldots \rho^{a_{n}} (\underline
{x}^{n})> = { \int [d \rho] \rho^{a_1} (\underline {x}^1) \ldots
\rho^{a_{n}} (\underline {x}^{n})\exp \left( - \int d^2 {x} {{\rho ^2}
( \underline{x}) \over {2{\mu}^2 (\underline {x}) }} \right) \over
{\int [d \rho] \exp \left( - \int d^2 {x} {{\rho ^2} ( \underline{x})
\over {2{\mu}^2 (\underline {x}) }} \right) }} ,
\label{conc1}
\end{eqnarray}
with 
\begin{eqnarray}
\mu^2 ( \underline{x}) = { 3 g^2 \over 2} {C_F \over N_c} { N \over
{\pi R^2}} \sqrt {1 - {{\underline {x}}^2 \over R^2}} = { 3 g^2 \over
2} {C_F \over N_c} { 1 \over {\pi a^2}} N^{1/3} \sqrt {1 -
{{\underline {x}}^2 \over R^2}}
\end{eqnarray}
in our model. Note that our $\mu^2$ goes as $N^{1/3}$, as
expected. The average on the left hand side is understood as $<\ldots>
= \prod_{k=1}^N \int { d^3 r_k \over {(4/3) \pi R^3}} { d^3 x_k d^3
x'_k \over {[(4/3)\pi a^3]^2}} (\alpha {\overline \alpha}| \ldots
|\beta {\overline \beta})$. For a spherical nucleus $\mu^2$ is a
function of $\underline {x}$.

      The most general observable in our system can be represented as
some functional of $\int K({x}' , \underline {x})\rho (\underline {x})
d^2 {x}$, namely the value of this observable for a given $\rho
(\underline {x})$ is
\begin{eqnarray}
{\cal O}_{\rho} = F \left( \int K({x}' , \underline {x}) \rho
  (\underline {x}) d^2 {x} \right),
\label{conc2}
\end{eqnarray}
where $F(f)$ is some functional and $K({x}' , \underline {x})$ is some
kernel independent of $\rho (\underline {x})$.  For instance, one can
consider Wilson loop: $W=<\mbox {Tr P} \exp \left( -ig \int d x \cdot
{\bf A} \right)>$. It's a functional of ${\bf A} (x)$. The field ${\bf
A} (x)$ can be represented as $\int K({x}' ,\underline {x}) \rho
(\underline {x}) d^2 {x} $. Then, using the definition of a
path-ordered integral, we can expand the Wilson loop in the powers of
${\bf A} (x)$:
\begin{eqnarray}
W = <\mbox {Tr} \prod_k [ 1 - i g d x_k \cdot {\bf A} (x_k) ]> = <
\mbox {Tr} \sum_n c_n {\bf A} (x_1) \ldots {\bf A} (x_n) >,
\end{eqnarray}
with some coefficients $c_n$. Writing ${\bf A} (x)$ as ${\bf A} (x) =
\sum_{a=1}^8 T^a \int K^a ({x} ,\underline {x}') \rho (\underline
{x}') d^2 {x}'$, we achieve
\begin{eqnarray}
W= \sum_n c_n \sum_{a_1 \ldots a_n =1}^8 \mbox {Tr} [T^{a_1} \ldots
T^{a_n}] \int d^2 x'_1 \ldots d^2 x'_n K^{a_1} ({x_1} ,\underline
{x}'_1) \ldots K^{a_n} ({x_n} ,\underline {x}'_n) \times \nonumber
\end{eqnarray}
\begin{eqnarray}
\times < \rho (\underline {x}'_1) \ldots \rho (\underline {x}'_n) > .
\end{eqnarray}
Now we can use (\ref{conc1}) , getting
\begin{eqnarray}
W = { \int [d \rho] \exp \left( - \int d^2 {x} {{\rho ^2} (
\underline{x}) \over {2{\mu}^2 (\underline {x}) }} \right) \mbox{Tr P}
\exp \left( -ig \int d x \cdot {\bf A} \right) \over {\int [d \rho]
\exp \left( - \int d^2 {x} {{\rho ^2} ( \underline{x}) \over {2{\mu}^2
(\underline {x}) }} \right) }} .
\end{eqnarray}
A similar treatment can be applied to any observable ( which can be
represented as (\ref{conc2})) to prove that
\begin{eqnarray}
<{\cal O}_{\rho}> = {\int [d \rho] \exp \left( - \int d^2 {x} {{\rho
^2} ( \underline{x}) \over {2{\mu}^2 (\underline {x}) }} \right) {\cal
O}_{\rho} \over {\int [d \rho] \exp \left( - \int d^2 {x} {{\rho ^2} (
\underline{x}) \over {2{\mu}^2 (\underline {x}) }} \right) }}.
\end{eqnarray}
This confirms the assumption made by McLerran and Venugopalan in their
model [1]\renewcommand{\thefootnote}{\fnsymbol{footnote}}
\footnote{I have recently learned (private communication from
L. McLerran to A. Mueller) that similar conclusions have been reached
by Professor L. McLerran and collaborators.}.

\section*{Acknowledgements}

 I want to thank my adviser Professor A.H. Mueller for his guidance
and advice. Without his support this work would not be possible.

   This research is sponsored in part by the US Department of Energy
under grant DE-FG02 94ER 40819.

\appendix
 
\section{}

From [3] the light-cone potential of a charge $e$ moving with a
velocity $v$ is
\begin{eqnarray}
A_{+} = { \sqrt 2 e \over { 4 \pi }} {1 \over {\sqrt {2 x_{-}^2 +
(1-v^2) {\underline {x}}^2}}}, \underline A = A_{-} = 0.
\end{eqnarray}
If we do a Fourier transform
\begin{eqnarray}
A_{\mu} (k) = {1 \over { (2 \pi)^2 }} \int dx_{+} dx_{-} d^2
\underline {x} e^{i k_{+} x_{-} + i k_{-} x_{+} - i {\underline {k}}
\cdot {\underline {x}}} A_{\mu} (x)
\end{eqnarray}
and take the limit $v \rightarrow 1 $, we get for the
ultrarelativistic particle:
\begin{eqnarray}
A_{+} (k) = {e \delta (k_{-}) \over {2 \pi {\underline {k}}^2 }} .
\label{xA1}
\end{eqnarray}
If we go back to the coordinate space, by doing an inverse Fourier
 transform of (\ref{xA1}), we end up with
\begin{eqnarray}
A_{+} (x) = - {e \over {2 \pi}} \delta (x_{-}) \ln (|\underline {x}|
\lambda),
\end{eqnarray} 
which is different form the $v \rightarrow 1 $ limit of the original
expression by a gauge transformation.

\section{}

 Defining
\begin{eqnarray}
f_i (\underline x, x_{-}) = { g \over {2 \pi }} \sum_{a=1}^8
(T_i^a)S(\underline {x}, x_{-} ) T^a S^{-1} (\underline {x}, x_{-})
\end{eqnarray}
we can rewrite (\ref{clsol}) as 
\begin{eqnarray}
{\underline {A}}(\underline {x}, x_{-} ) = \sum_{i=1}^N \left( f_i
(\underline x, x_{-i}){{\underline{x}}-{\underline{x}}_i \over
|{\underline{x}}-{\underline{x}}_i|^2}\theta (x_{-} - x_{-i}) - f_i
(\underline x, x'_{-i}){{\underline{x}}-{\underline{x}'}_i \over
|{\underline{x}}-{\underline{x}'}_i|^2}\theta (x_{-} - x'_{-i})
\right).
\label{xB1}
\end{eqnarray}
Then the field strength is
\begin{eqnarray}
F_{+\perp} (\underline {x}, x_{-} ) = {\partial \over {\partial x_{-}
}} {\underline {A}}(\underline {x}, x_{-} ) = \sum_{i=1}^N \left( f_i
(\underline x, x_{-i}){{\underline{x}}-{\underline{x}}_i \over
|{\underline{x}}-{\underline{x}}_i|^2}\delta (x_{-} - x_{-i})
\right. \nonumber\\ \left. - f_i (\underline x,
x'_{-i}){{\underline{x}}-{\underline{x}'}_i \over
|{\underline{x}}-{\underline{x}'}_i|^2}\delta (x_{-} - x'_{-i})
\right).
\label{xB2}
\end{eqnarray}
Substituting (\ref{xB1}) and (\ref{xB2}) into (\ref{dens1}) we get
\begin{eqnarray}
\rho (\underline {x}) = \sum_{i=1}^N [ f_i (\underline x, x_{-i}) 2
\pi \delta(\underline x - \underline x_{i}) - f_i (\underline x,
x'_{-i}) 2 \pi \delta (\underline x - \underline {x}'_{i}) ]
\nonumber\\ + \sum_{i=1}^N \left( \nabla f_i (\underline x, x_{-i})
{{\underline{x}}-{\underline{x}}_i \over
|{\underline{x}}-{\underline{x}}_i|^2} - \nabla f_i (\underline x,
x'_{-i}) {{\underline{x}}-{\underline{x}'}_i \over
|{\underline{x}}-{\underline{x}'}_i|^2} \right) \nonumber\\ - i g
\sum_{i,j=1}^N \left( [f_i (\underline x, x_{-i}),f_j (\underline x,
x_{-j})] {{\underline{x}}-{\underline{x}}_i \over
|{\underline{x}}-{\underline{x}}_i|^2}
{{\underline{x}}-{\underline{x}}_j \over
|{\underline{x}}-{\underline{x}}_j|^2} \theta (x_{-j} - x_{-i} )
\right. \nonumber\\ - [f_i (\underline {x}, x_{-i}),f_j (\underline x,
x'_{-j})] {{\underline{x}}-{\underline{x}}_i \over
|{\underline{x}}-{\underline{x}}_i|^2}
{{\underline{x}}-{\underline{x}'}_j \over
|{\underline{x}}-{\underline{x}'}_j|^2} \theta (x'_{-j} - x_{-i} )
\nonumber\\ - [f_i (\underline x, x'_{-i}),f_j (\underline x, x_{-j})]
{{\underline{x}}-{\underline{x}'}_i \over
|{\underline{x}}-{\underline{x}'}_i|^2}
{{\underline{x}}-{\underline{x}}_j \over
|{\underline{x}}-{\underline{x}}_j|^2} \theta (x_{-j} - x'_{-i} )
\nonumber\\ \left. + [f_i (\underline x, x'_{-i}),f_j (\underline x,
x'_{-j})] {{\underline{x}}-{\underline{x}'}_i \over
|{\underline{x}}-{\underline{x}'}_i|^2}
{{\underline{x}}-{\underline{x}'}_j \over
|{\underline{x}}-{\underline{x}'}_j|^2} \theta (x'_{-j} - x'_{-i} )
\right).
\label{xB3}
\end{eqnarray}
A straightforward calculation yields
\begin{eqnarray}
 \nabla f_i (\underline x, x_{-i}) = { i g^3 \over {(2 \pi)^2 }}
\sum_{a=1}^8 \sum_{b=1}^8 \sum_{j=1}^N (T^a_i) (T^b_j) \{
{{\underline{x}}-{\underline{x}}_j \over
|{\underline{x}}-{\underline{x}}_j|^2}\theta(x_{-i}-x_{-j}) \nonumber
\end{eqnarray}
\begin{eqnarray}
\times [ S(\underline x, x_{-j}) T^b W(\underline x, x_{-i} , x_{-j})
T^a S^{-1} (\underline x, x_{-i}) - S(\underline x, x_{-i}) T^a
W(\underline x, x_{-j} , x_{-i}) \nonumber
\end{eqnarray}
\begin{eqnarray}
\times T^b S^{-1} (\underline x, x_{-j}) ] -
{{\underline{x}}-{\underline{x}'}_j \over
|{\underline{x}}-{\underline{x}'}_j|^2}\theta (x_{-i} - x'_{-j})[
S(\underline x, x'_{-j}) T^b W(\underline x, x_{-i} , x'_{-j}) T^a
\nonumber
\end{eqnarray}
\begin{eqnarray}
\times S^{-1} (\underline x, x_{-i}) - S(\underline x, x_{-i}) T^a
W(\underline x, x'_{-j} , x_{-i}) T^b S^{-1} (\underline x, x'_{-j})]
\}
\end{eqnarray}
and
\begin{eqnarray}
[f_i (\underline x, x_{-i}),f_j (\underline x, x_{-j})] = { g^2 \over
{(2 \pi)^2 }} \sum_{a=1}^8 \sum_{b=1}^8 (T^a_i) (T^b_j) [
S(\underline x, x_{-i}) T^a W(\underline x, x_{-j} , x_{-i}) \nonumber
\end{eqnarray}
\begin{eqnarray}
\times T^b S^{-1} (\underline x, x_{-j}) - S(\underline x, x_{-j}) T^b
W(\underline x, x_{-i} , x_{-j}) T^a S^{-1} (\underline x, x_{-i}) ] ,
\end{eqnarray}
where 
\begin{eqnarray}
W(\underline x, x_{-i} , x_{-j}) = \mbox {P} \exp \left( - i g
\int_{x_{-j}}^{x_{-i}} dx_{-} A'_{+} (\underline x, x_{-}) \right).
\end{eqnarray}
Plugging this back into (\ref{xB3}) we end up with
\begin{eqnarray}
\rho (\underline {x}) = \sum_{i=1}^N [ f_i (\underline x, x_{-i}) 2
\pi \delta(\underline x - \underline x_{i}) - f_i (\underline x,
x'_{-i}) 2 \pi \delta (\underline x - \underline {x}'_{i}) ],
\end{eqnarray}
which is equivalent to (\ref{dens2}).

\section{}

Let's prove that the three-densities correlation function is
zero. Then the proof for an arbitrary correlation function of an odd
number of densities will become obvious. Similar to (\ref{corr1}) we
write
\begin{eqnarray}
<\rho^a (\underline {x})\rho^b (\underline {y})\rho^c (\underline
{z})> = { g^3 } \prod_{l=1}^N \int { d^3 r_l \over
{(4/3) \pi R^3}} { d^3 x_l d^3 x'_l \over {[(4/3) \pi a^3]^2}} (\alpha
{\overline \alpha}| \sum_{a',b',c'=1}^8 \sum_{i,j,k=1}^N (T^{a'}_i)
(T^{b'}_j) (T^{c'}_k) \nonumber
\end{eqnarray}
\begin{eqnarray}
\times \left( 2 \mbox{Tr} [ T^a S(\underline {x}_i, x_{-i} ) T^{a'}
S^{-1} (\underline {x}_i, x_{-i})] \delta (\underline {x} - \underline
{x}_i) - 2 \mbox{Tr} [ T^a S(\underline {x}'_i, x'_{-i} ) T^{a'}
S^{-1} (\underline {x}'_i, x'_{-i})] \delta (\underline {x} -
\underline {x}'_i) \right) \nonumber
\end{eqnarray}
\begin{eqnarray}
\times \left( 2 \mbox{Tr} [ T^b S(\underline {x}_j, x_{-j} ) T^{b'}
S^{-1} (\underline {x}_j, x_{-j})] \delta (\underline {y} - \underline
{x}_j) - 2 \mbox{Tr} [ T^b S(\underline {x}'_j, x'_{-j} ) T^{b'}
S^{-1} (\underline {x}'_j, x'_{-j})] \delta (\underline {y} -
\underline {x}'_j) \right) \nonumber
\end{eqnarray}
\begin{eqnarray}
\times \left( 2 \mbox{Tr} [ T^c S(\underline {x}_k, x_{-k} ) T^{c'}
S^{-1} (\underline {x}_k, x_{-k})] \delta (\underline {z} - \underline
{x}_k) - 2 \mbox{Tr} [ T^c S(\underline {x}'_k, x'_{-k} ) T^{c'}
S^{-1} (\underline {x}'_k, x'_{-k})] \delta (\underline {z} -
\underline {x}'_k) \right)|\beta {\overline \beta}).
\end{eqnarray}
After dropping the ``last'' nucleon and averaging over the colors of
this nucleon, we get
\begin{eqnarray}
<\rho^a (\underline {x})\rho^b (\underline {y})\rho^c (\underline
{z})> = { 8 g^3 } \prod_{l=1}^N \int { d^3 r_l
\over {(4/3) \pi R^3}} { d^3 x_l d^3 x'_l \over {[(4/3) \pi a^3]^2}}
\nonumber
\end{eqnarray}
\begin{eqnarray}
\times (\alpha {\overline \alpha}| \sum_{a',b',c'=1}^8 \sum_{i=1}^N
\mbox{Tr} [(T^{a'}_i) (T^{b'}_i) (T^{c'}_i)] \mbox{Tr} [ T^a S_x
T^{a'} S_x^{-1}] \mbox{Tr} [ T^b S_y T^{b'} S_y^{-1}] \mbox{Tr} [ T^c
S_z T^{c'} S_z^{-1}] |\beta {\overline \beta}) \nonumber
\end{eqnarray}
\begin{eqnarray}
\times [ \delta (\underline {x} - \underline {x}_i) - \delta
(\underline {x} - \underline {x}'_i) ] [ \delta (\underline {y} -
\underline {x}_i) - \delta (\underline {y} - \underline {x}'_i) ] [
\delta (\underline {z} - \underline {x}_i) - \delta (\underline {z} -
\underline {x}'_i) ] ,
\end{eqnarray}
where $S_x=S(\underline {x}, x_{-i} )=S(\underline {x}, x'_{-i} )$
(after dropping the ``last'' nucleon) and independent of $\underline
{x}_i$. The product of three brackets with delta-functions integrated
over $\underline {x}_i$ and ${\underline {x}}'_i$ obviously gives
zero. So, $ <\rho^a (\underline {x})\rho^b (\underline {y})\rho^c
(\underline {z})> = 0 $ , as advertised.

     Similar techniques can be applied to an arbitrary odd number of
densities to show that their correlation function is zero.

\end{document}